\documentclass{JHEP3}

\usepackage{epsfig}

\newcommand{\be}{\begin{equation}}
\newcommand{\ee}{\end{equation}}

\title{Large $N$ QCD in two dimensions
with a baryonic chemical potential}
\author{ Richard Galvez, Ari Hietanen and Rajamani Narayanan
\\Department of Physics, Florida International University,
Miami, FL 33199.}

\abstract{We consider large $N$ gauge theory on a 
two dimensional lattice in the presence of 
a baryonic chemical potential. We work with one copy
of na\"ive fermion and
argue that reduction holds even in the presence of
a chemical potential. Analytical arguments
supported by numerical studies 
show that there is no phase transition as a function
of the baryonic chemical potential.
}


\preprint{}

\begin{document}
\section{Introduction}
The 't Hooft model~\cite{'tHooft:1974hx}, namely, large $N$ gauge
theory with finite number of fermion flavors in two dimensions, has
several similarities with four dimensional QCD.  The particle spectrum
has an infinite tower of mesons~\cite{'tHooft:1974hx}, and there is a
massless meson in the chiral limit.  Chiral symmetry is broken in the
$N\to\infty$ limit even though it is a two dimensional
model. Confinement is simply realized since Wilson loops obey an exact
area law~\cite{Migdal:1975zg,Rusakov:1990rs}.

Eguchi-Kawai reduction~\cite{Eguchi:1982nm}
 holds in two dimensions~\cite{Bhanot:1982sh}
and the model can be reduced to a single site on the lattice.
The two $Z_N$ symmetries 
($U(1)$ in the $N\to\infty$ limit) associated with the Polyakov loops
in the two directions remain unbroken
as one approaches the continuum limit.
Therefore, the model does not depend on the spatial extent
or the temperature and it is always in the confined phase.
Since finite number of fermion flavors do not affect the gauge field
dynamics in the $N\to\infty$ limit~\cite{'tHooft:1973jz},
the chiral condensate can be
computed in the ``quenched approximation''
and is independent of the temperature.

We will study the role of the baryonic chemical potential
in the Euclidean formalism of lattice large $N$ QCD. 
The $Z_N$ symmetries associated with Polyakov loops will
play a central role in our discussion and we will use
a generalized partition function 
to obtain a result for the quark number density
with a single copy (four degenerate flavors) of
na\"ive fermion. The main result of this paper is the
independence of the physics on the quark chemical potential.
We provide analytical arguments and support it with numerical
calculations.

\section{The quark number density}

We consider the generalized
partition function of lattice QCD on a
 two dimensional $L_1\times L_2$ lattice
with one copy of na\"ive fermion
defined by
\be
Z(\phi,\chi) = 
\int \prod_x [dU_1(x)] [dU_2(x)] 
e^{2bN\sum_p {\rm Re\ Tr}U_p} 
\det {\cal M} \left(T_1,T_2,m_q,\phi,\chi\right),
\label{gpf}
\ee
where $U_\mu(x)$ are SU(N) matrices on the links connecting
site $x$ to $x+\hat\mu$ for $\mu=1,2$.
$[dU_1(x)]$ and $[dU_2(x)]$ are the standard Haar measures on SU(N).
$b$ is the lattice gauge coupling
which is kept constant as $N$ is taken to infinity.
The continuum limit is obtained by taking $b\to\infty$
and the lattice spacing can be set to 
$\frac{1}{\sqrt{b}}.$\footnote{This amounts to fixing the string
tension at $\frac{1}{4}$ as one approaches the 
continuum limit~\cite{Gross:1980he}.}
$U_p$ is
the standard plaquette operator.
$\phi$ and $\chi$ are two independent complex variables.
We will assume that $N$ is odd so that the baryons are
made up of an odd number of quarks. The na\"ive Dirac operator is
\be
{\cal M}\left(T_1,T_2,m_q,\phi,\chi \right)
= 2m_q + \sigma_1 \left(\phi T_1 
- \phi^{-1} T_1^\dagger \right)
+\sigma_2\left (\chi T_2 
-\chi^{-1} T_2^\dagger\right)\label{fermion},
\ee 
where
$T_\mu$ are unitary operators given by
\be
\left[T_\mu \psi\right](x) = U_\mu(x)\psi(x+\hat\mu).
\ee
${\cal M}$ is a matrix with a linear extent of 
$M=2NV$ with $V=L_1L_2$.

It is clear that the result is a finite polynomial of the
form
\be
Z(\phi,\chi) = \sum_{k_2=-M}^{M} \sum_{k_1=-M+|k_2|}^{M-|k_2|}
a_{k_1,k_2} \phi^{k_1}\chi^{k_2}.\label{poly}
\ee
Since the result is a finite polynomial, one can obtain all
the coefficients by setting $\phi=e^{ip}$ and $\chi=e^{i\omega}$
with $p,\omega\in [-\pi,\pi]$ and using fourier transforms.
This is by no means a new idea~\cite{Lombardo:2006yc}
nor does it avoid the sign problem since it involves
fourier transforms of measured quantities.

The fermion determinant appearing in the definition
of $Z(e^{ip},e^{i\omega})$ is positive definite and
it does not affect the gauge dynamics since the effect
of $p$ and $\omega$ is simply to make the fermions
interact with a constant external $U(1)$ field on top of the
$SU(N)$ gauge field. Therefore, we can invoke 
Eguchi-Kawai~\cite{Eguchi:1982nm}
reduction and write $Z(e^{ip},e^{i\omega})$ as
\begin{eqnarray}
Z(e^{ip},e^{i\omega}) = 
\int && \prod_x [dU_1] [dU_2] 
e^{2bN {\rm Re\ Tr}U_1U_2U_1^\dagger U_2^\dagger} \cr
&& \prod_{k_1=0}^{L_1-1}
\prod_{k_2=0}^{L_2-1}
\det {\cal M} \left(U_1,U_2,m_q,
e^{i\left(p+\frac{2\pi k_1}{L_1}\right)},
e^{i\left(\omega+\frac{2\pi k_2}{L_2}\right)}\right).
\end{eqnarray}
We expect factorization of observables in the $N\to\infty$ limit
and therefore, we can write
\begin{eqnarray}
Z_f(e^{ip},e^{i\omega}) &=&
\frac{Z(e^{ip},e^{i\omega})}{\int \prod_x [dU_1] [dU_2] 
e^{2bN {\rm Re\ Tr}U_1U_2U_1^\dagger U_2^\dagger} }\cr
&=& \prod_{k_1=0}^{L_1-1}
\prod_{k_2=0}^{L_2-1} 
{\cal D}_f\left(e^{i\left(p+\frac{2\pi k_1}{L_1}\right)},
e^{i\left(\omega+\frac{2\pi k_2}{L_2}\right)}\right)
\label{factorize}
\end{eqnarray}
where
\be
{\cal D}_f\left(\phi,\chi\right)
=
\frac{
\int \prod_x [dU_1] [dU_2] 
e^{2bN {\rm Re\ Tr}U_1U_2U_1^\dagger U_2^\dagger} 
\det {\cal M} \left(U_1,U_2,m_q,\phi,\chi\right)}
{\int \prod_x [dU_1] [dU_2] 
e^{2bN {\rm Re\ Tr}U_1U_2U_1^\dagger U_2^\dagger} }
\label{ekmodel}
\ee

The gauge action and the Haar measure 
in (\ref{ekmodel}) are invariant
under $U_1\to e^{i\frac{2\pi}{N}}U_1$
and $U_2\to e^{i\frac{2\pi}{N}}U_2$.
Therefore, 
\be
{\cal D}_f(\phi,\chi) = {\cal D}_f\left(e^{i\frac{2\pi}{N}}\phi,
\chi\right) =
{\cal D}_f\left(\phi,e^{i\frac{2\pi}{N}}\chi
\right) 
\ \ \ \forall\ \ \ \phi,\chi
.\label{zn}
\ee
It follows from (\ref{fermion}) that
\be
\sigma_1
{\cal M}(U_1,U_2,m_q,\phi,\chi)
\sigma_1
={\cal M}(U_1,U_2,m_q,\phi,-\chi)
\label{flip1}
\ee
and
\be
\sigma_2
{\cal M}(U_1,U_2,m_q,\phi,\chi)
\sigma_2
={\cal M}(U_1,U_2,m_q,-\phi,\chi).
\label{flip2}
\ee

Equations (\ref{zn}), 
(\ref{flip1}) and (\ref{flip2})
 along with (\ref{poly}) restricted to $V=1$
results in a 
polynomial of the form
\be
{\cal D}_f(\phi,\chi) =
a_{0,0} + a_{2N,0} \phi^{2N} + a_{-2N,0}\phi^{-2N}
+a_{0,2N} \chi^{2N} + a_{0,-2N} \chi^{-2N}.
\label{zw}
\ee

Since $a_{\pm 2N,0}$ and $a_{0, \pm 2N}$ correspond
to the highest or lowest term in the polynomial
in the respective variables, it follows that
\be
a_{2N,0}=\langle \det \sigma_1 U_1 \rangle ;\ \ 
a_{-2N,0}=\langle \det \sigma_1 U_1^\dagger \rangle ;\ \ 
a_{0,2N}=\langle \det \sigma_2 U_2 \rangle ;\ \ 
a_{0,-2N}=\langle \det \sigma_2 U_2^\dagger \rangle ,
\ee
and hence all these coefficients are $-1$ for odd $N$.

Therefore, we finally have,
\be
{\cal D}_f(\phi,\chi) = a_{0,0} 
- \left(\phi^{2N} + \phi^{-2N}\right)
- \left(\chi^{2N} + \chi^{-2N}\right)
\label{ekdf}
\ee

The number density, $\rho(\mu)$,
at zero temperature and
infinite spatial extent ($L_1,L_2\to\infty$) at a
fixed chemical potential, $\mu$, 
is given by
\be
\rho(\mu) = \lim_{L_1,L_2\to\infty}\frac{1}{NL_1L_2}
\frac{\partial}{\partial\mu} \log 
Z(1,e^\mu)
\label{nden}
\ee
Using (\ref{factorize}) and (\ref{ekdf}), we get
\be
\rho(\mu) = 
\frac{1}{2\pi} \int_{-\pi}^\pi dp\  n(p,\mu)
\ee
where
\be
n(p,\mu)=
\frac{1}{2\pi} \int_{-\pi}^\pi d\omega
\frac{-4i\sin 2N(\omega -i \mu)}{a_{0,0} 
- 2\cos 2Np
- 2\cos 2N(\omega-i\mu)\label{ferdet}
}.
\label{numden}
\ee
This result for the number density with
an explicit integral over the momenta, $p$, and
the Matsubara frequencies, $\omega$, for the
interacting theory in the large $N$ limit is
an example of the quenched momentum 
prescription~\cite{Gross:1982at}.

Let us set
\be
a_{0,0} = 4M^2 + 4.
\ee
We can rewrite (\ref{numden}) as
\be
n(p,\mu) = \frac{4N}{i\pi} \int_0^{\frac{\pi}{N}} d\omega
\frac {\sin 2N\left(\omega - i\mu\right)}
{4M^2 +2 + 4\sin^2 Np - 2\cos 2N(\omega-i\mu)}.
\label{npmu}
\ee
With $z=e^{2Ni\omega+2N\mu}$, the above integral becomes
a contour integral over a circle of radius
$e^{2N\mu}$ centered at the origin and
\begin{eqnarray}
n(p,\mu)&=&\frac{1}{i\pi}\oint \frac{z^2-1}
{z\left[z^2+1-2\left(1+2\left\{M^2+\sin^2 Np\right\} 
\right)z\right]}dz
\cr &=&\cases{ 
0 & 
if $\mu < 
\frac{1}{N}\sinh^{-1} \sqrt{\sin^2Np+M^2}$
\cr
2 & if $\mu > 
\frac{1}{N}\sinh^{-1} \sqrt{\sin^2Np+M^2}$
\cr}.
\label{inter}
\end{eqnarray}
Writing, $\mu=\frac{\bar\mu}{\sqrt{b}}$ with $\bar\mu$ being
the physical chemical potential, we see that
the critical chemical potential, $\bar\mu_c$, is
given by
\be
\bar\mu_c = \lim_{b\to\infty} \sqrt{b} \lim_{N\to\infty}
\frac{1}{N}\sinh^{-1} \sqrt{\sin^2Np+M^2}.
\label{critchem}
\ee

By comparing this result with that for free quarks, namely,
\begin{eqnarray}
n(p,\mu)&=&\frac{1}{i\pi}\oint \frac{z^2-1}
{z\left[z^2+1-2\left(1+2\left\{m_q^2+\sin^2p\right\}\right)z\right]}dz
\cr
&=&\cases{ 
0 & if $\mu < \sinh^{-1}\sqrt{\sin^2 p +m_q^2}$ \cr
2 & if $\mu > \sinh^{-1}\sqrt{\sin^2 p +m_q^2}$ \cr}
\label{naive}
\end{eqnarray}
we see that
the effect of the interactions is to scale
$\mu$ and $p$ by a factor of $N$ 
($Np$ and $N\mu$ can be viewed as the
baryonic mometum and baryonic chemical potential)
and replace the quark mass,
$m_q$,
by $M$ defined as
\be
M^2=\frac{1}{4} {\cal D}_f(1,1) = \frac{1}{4}
\left\langle \det\left [  2m_q + \sigma_1 \left(U_1 
- U_1^\dagger\right)
+ \sigma_2 \left(U_2 - U_2^\dagger \right)
\right]\right\rangle.
\label{deqn}
\ee

\section{The physics of $M$}\label{mphys}

It is clear from (\ref{deqn}) that $M=m_q$ for $U_1=U_2=1$.
It is also clear that
$M$ will behave like $e^{\alpha N}$ 
for large $N$ and $\alpha$ will depend
on $b$ and $m_q$. Following 't Hooft~\cite{'tHooft:1974hx}, 
we will keep $\gamma = 2\pi b m_q^2$ fixed as we take the
continuum limit, $b\to\infty$. 

We can compute the ``tree level'' contribution 
to $M$ by setting
\be
  U_1 = e^{i{\rm diag}(u_1,u_2,\dots,u_{N})}
 ~~ U_2 = e^{i{\rm diag}(v_1,v_2,\dots,v_{N})},
\label{treeg}
\ee
where $u_i, v_i \in [-\pi,\pi]$ and $\sum_{i=1}^{N} u_i = 
\sum_{i=1}^{N} v_i = 0$
since we are in the confined phase.
Setting
$m_q=0$, 
we have
\be
M^2
= I_N(0,0),
\ee
where $I_N(p,q)$ is defined in Appendix~\ref{intI} with
\be
f(u,v) = 4\left( \sin^2 u +\sin^2 v\right) =
4 - \left( e^{2i u} + e^{-2iu} + e^{2i v} + e^{-2i v}\right).
\label{naivesingle}
\ee
Therefore,
\be
M^2
= 4^N + 4(-1)^N,
\ee
and the critical chemical potential
$\bar\mu_c$ in (\ref{critchem})
will approach infinity as $\ln(2)\sqrt{b}$ when
$b\to\infty$.
This is the main result of our paper.

Following the steps similar to the ones from (\ref{nden})
to (\ref{inter}), we can show that the 
chiral condensate, $\chi(\mu)$,
defined as 
\be
\chi(\mu) = \lim_{m_q\to 0}
\lim_{L_1,L_2\to\infty}\frac{1}{NL_1L_2}
\frac{\partial}{\partial m_q} \log 
Z(1,e^\mu)
\label{chiral}
\ee
reduces to
\be
\chi(\mu) = \cases{
\lim_{m_q\to 0} 2 \frac{d\alpha}{dm_q} &
if $\mu < 
\frac{1}{N}\sinh^{-1} \sqrt{\sin^2Np+M^2}$
\cr
0 & if $\mu > 
\frac{1}{N}\sinh^{-1} \sqrt{\sin^2Np+M^2}$
\cr}.
\ee
Since the chiral condensate should behave as
\be
\chi(0)=\frac{\Sigma}{\sqrt{b}},
\ee
we expect 
$\alpha$ to take the form
\be
\alpha = \ln (2) + \frac{\mu_0}{\sqrt{b}} +  
\frac{\Sigma}{2\sqrt{2\pi} b} \sqrt{\gamma}
+\cdots \label{aleqn}
\ee
as we approach the continuum limit for large $N$.

\section{Numerical estimate of $\alpha$}\label{numerics}

We support our argument in Section~\ref{mphys}
by performing
a numerical estimate of $\alpha$. 
The main purpose is to convincingly show the presence
of $\ln(2)$ in (\ref{aleqn}). 
Our numerical procedure has the following steps:
\begin{enumerate}
\item We fix $N$, $b$ and $\gamma$ and numerically
evaluate the right hand side of (\ref{deqn}). 
\item We plot $\frac{1}{2}\ln {\cal D}_f(1,1) -N\ln 2$ 
as a function of $N$
at a fixed $b$ and $N$ to extract the large $N$ limit of
$\left(\alpha-\ln(2)\right)$. 
\item Based on (\ref{aleqn}),  we fit
\be
\alpha - \ln(2) = C_0(b) + C_1(b)\sqrt{\gamma} + O(\gamma)
\ee
and extract $C_0(b)$ and $C_1(b)$.
\item We fit $C_0(b)$ to 
\be
C_0 = z_0+ \frac{\mu_0}{\sqrt{b}} 
+ \frac{r_0}{b} + O(b^{-3/2})
\label{c0eqn}
\ee
with the aim of showing that $z_0$ is consistent with zero.
\item We fit $C_1(b)$ to 
\be
C_1 = z_1 + \frac{\mu_1}{\sqrt{b}} 
+ \frac{\Sigma}{2\sqrt{2\pi}b} + O(b^{-3/2})
\label{c1eqn}
\ee
with the aim of showing that $z_1$ and $\mu_1$
are consistent with zero.
\end{enumerate}

Our primary aim, as mentioned
above, is to show that $z_0$, $z_1$ and $\mu_1$ are consistent
with zero. The gauge action was simulated using
a Hybrid Monte Carlo algorithm since it
performs better than a Metropolis algorithm
based on SU(2) subgroups.
The simulations were performed for all odd values of $N$
in the range $\{11,39\}$. 
For each $N$ we used 12
values of $b=3,4,5,6,7,8,9,14,20,40,60,100$. 
We focused on going to smaller values of $b$ in order
to get close to the tree level result in Section~\ref{mphys}.
The HMC time step ranged from $0.02$ (at small $b$) to $0.004$ 
(at large
$b$). The number of steps in a trajectory was chosen so that the
length of the trajectory was $1.0$. 
The fermion determinant on each configuration was obtained
by performing an average over $N^2$ $Z_N$ transformed
gauge field configurations: $\left(
e^{i\frac{2\pi k_1}{N}}U_1,e^{i\frac{2\pi k_2}{N}}U_2\right)$;
$0\le k_1,k_2 < N$. This numerically enforces (\ref{zn}).
We computed the fermion determinant for
$21$ different values of masses in the range 
$\gamma=[0,4]$
on every configuration.

A sample plot 
at $b=3,8,20,100$ and $\gamma=3$ shown
in Fig.~\ref{fig1} illustrates the extraction of 
$\left(\alpha - \ln(2)\right)$. It is clear
from the slopes in Fig.~\ref{fig1}
that $\ln(2)$  dominates the value of $\alpha$ in our
range of $b$. Yet, we have sufficient statistics to
extract $\left(\alpha - \ln(2)\right)$
with reasonable accuracy even at the
weakest coupling.
\EPSFIGURE
{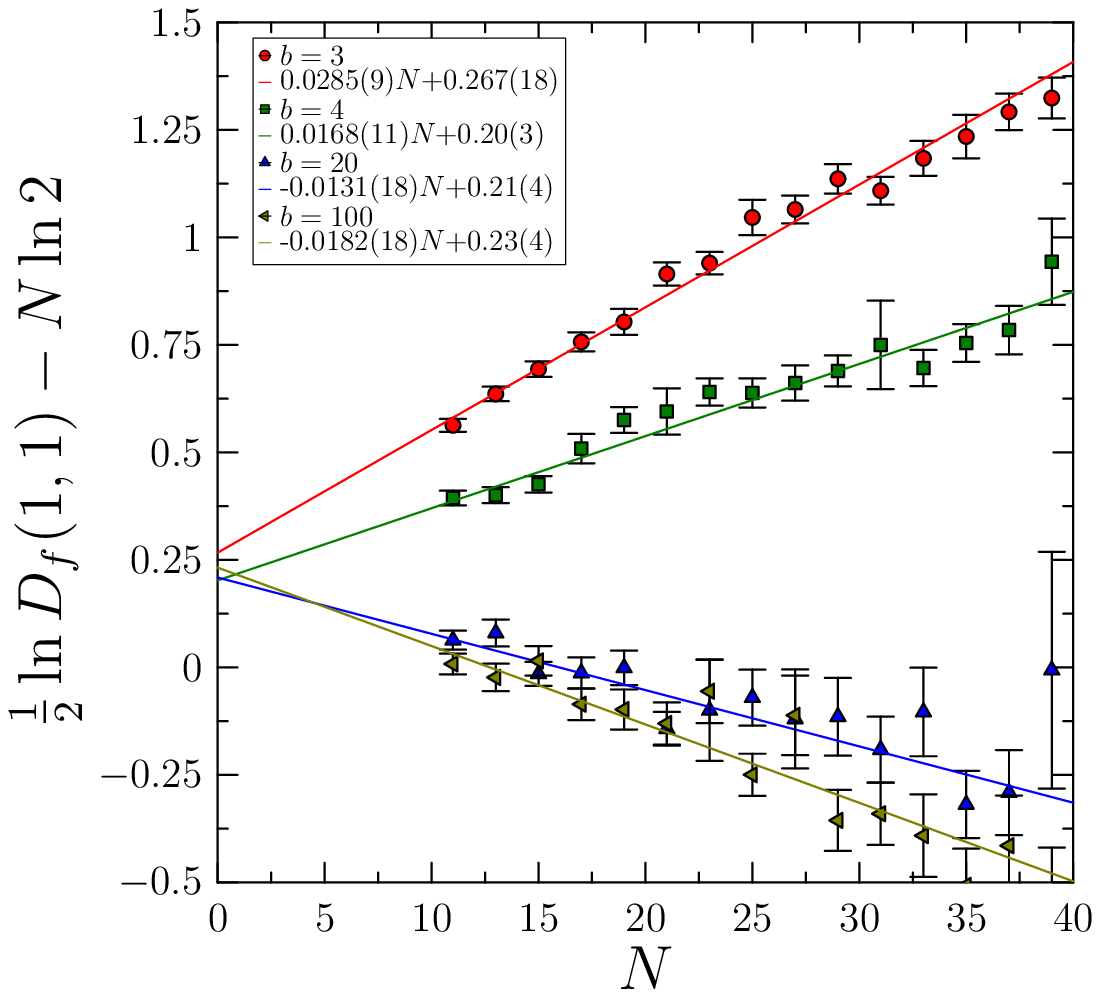, width=0.8\textwidth,clip=true}
{
Plot of $\frac{1}{2}\ln {\cal D}_f(1,1)-N\ln 2$ as a function of $N$
at $\gamma=3$.
\label{fig1}}

A sample plot at $b=3,8,20,100$ shown in Fig.~\ref{fig2}
illustrates the extraction of $C_0(b)$ and $C_1(b)$.
The shaded data points with $\gamma\ge 1$
were used to perform
the fit. Deviation from the linear behavior for $\gamma < 1$
is a consequence of finite $N$ effects. $\gamma =1$
is roughly the transition point between light and heavy
quarks~\cite{Kiskis:2002gr} and therefore we do not
expect to obtain a good estimate of $\Sigma$.
One should note that the errors in the intercept, $C_0(b)$,
are larger than the errors
in the slope, $C_1(b)$, and this is a consequence of
subtracting the dominant $\ln (2)$ term.
\EPSFIGURE
{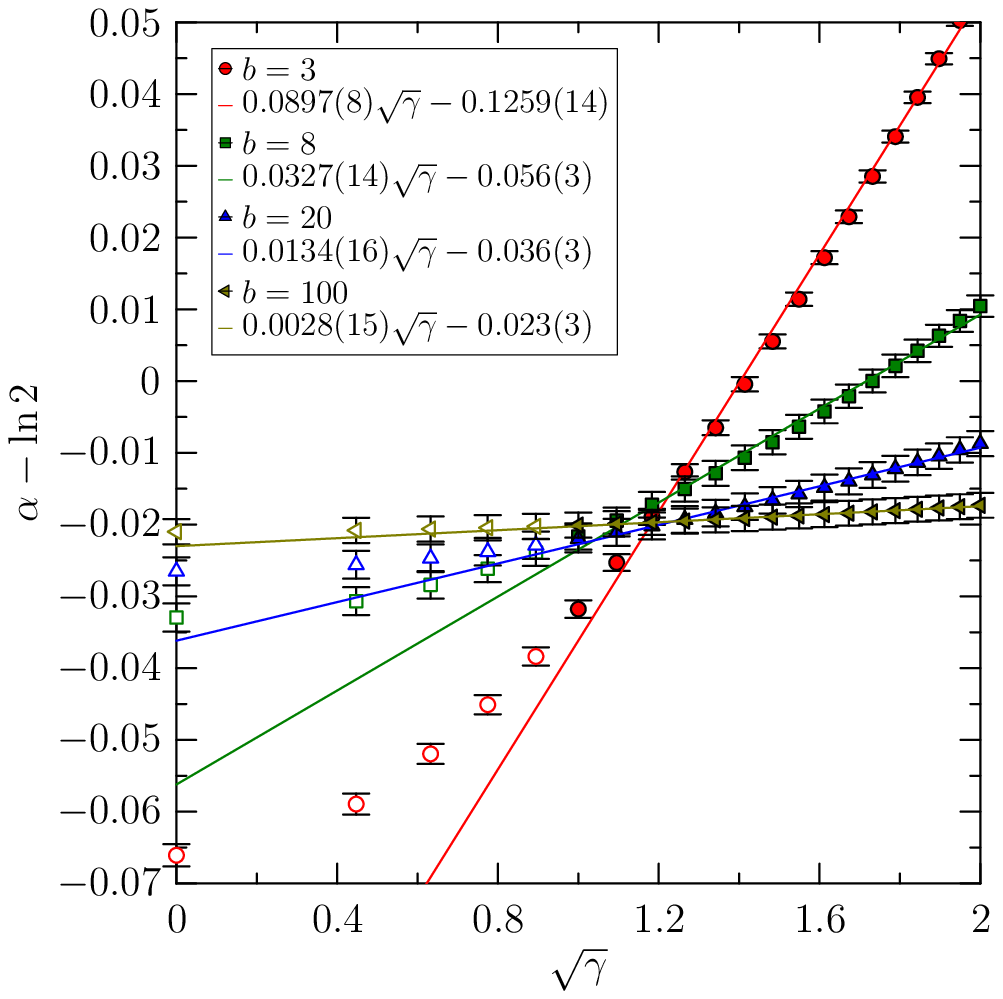, width=0.8\textwidth,clip=true}
{
Plot of $\alpha$ as a function of $\sqrt{\gamma}$.
\label{fig2}}

The plot of $C_0(b)$ vs $\frac{1}{\sqrt{b}}$ is shown in
Fig.~\ref{fig3}. The first observation is that the whole
range of the plot is less than $0.2\ln (2)$. The second
observation is a clear sign that $C_0(b)$ tends to zero
as $b\to\infty$. The fit of $C_0(b)$ to (\ref{c0eqn}),
shows that $z_0$ is zero within the expected accuracy
of this quantity. The other two fits, one with $z_0$ forced
to zero and the other with $\mu_0$ forced to zero, 
show a larger fluctuation in the non-zero coefficients
indicating that we do not have enough data to extract $\mu_0$ or
$r_0$.
\EPSFIGURE
{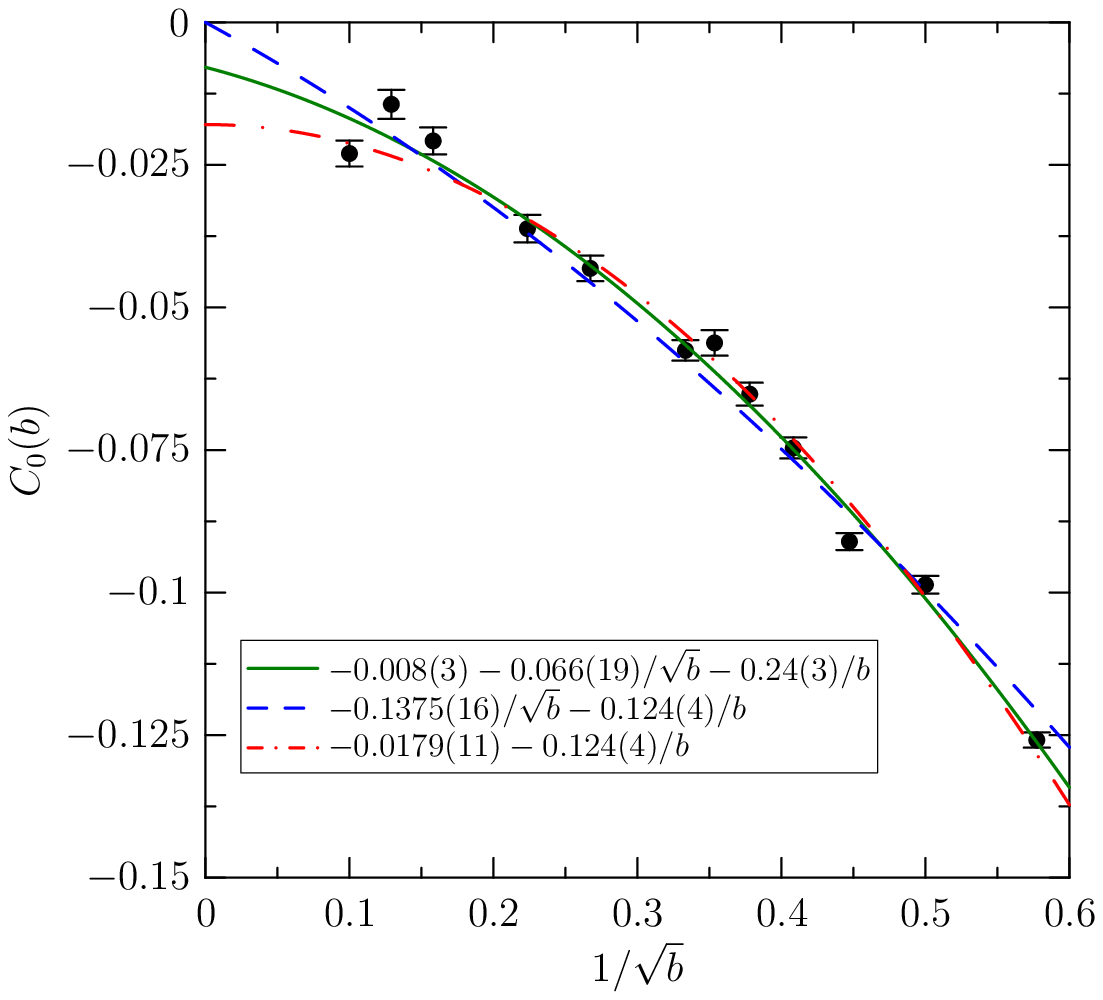, width=0.8\textwidth,clip=true}
{
Plot of $C_0(b)$ as a function of $b$.
\label{fig3}}

The plot of $C_1(b)$ vs $\frac{1}{\sqrt{b}}$ is shown in
Fig.~\ref{fig4}. 
A fit to the form in (\ref{c1eqn}) is
shown with solid lines. The fit is consistent with $z_1=0$
and $\mu_1=0$. To further emphasize this point, a fit
with $z_1$ forced to zero shows that $\mu_1$ is still
consistent with zero. Finally, a comparison of these
two fits to a fit with only the $\frac{1}{b}$ term
shows that the coefficient of the non-zero term 
remains unchanged in the three different cases.
Since na\"ive fermions in two dimensions describe
four flavors of fermions, the condensate for a single
flavor is estimated as $\frac{\Sigma}{4} = 0.34(3)$.
This estimate is significantly higher than the 
analytical result~\cite{Kiskis:2002gr} but this is
to be expected since we only used $\gamma > 1$ in our
fits.  
\EPSFIGURE
{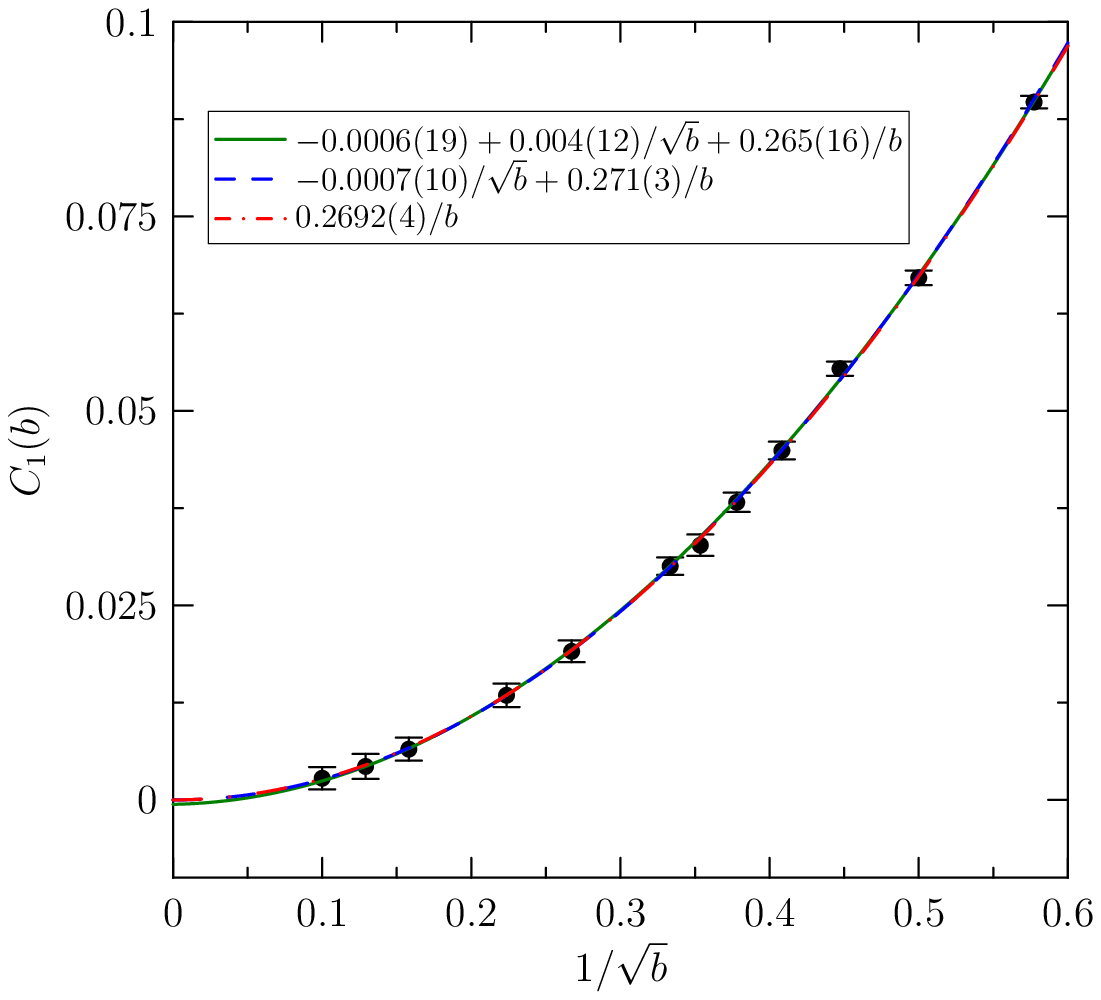, width=0.8\textwidth,clip=true}
{
Plot of $C_1(b)$ as a function of $b$.
\label{fig4}}

\section{Discussion}

The analysis in section~\ref{mphys} shows that
the absence of a critical chemical potential in two dimensional
large $N$ gauge theory can be seen at the tree level.
Numerical analysis presented in section~\ref{numerics}
supports the tree level argument. The basic result is
that the critical chemical potential behaves as $(\ln 2)\sqrt{b}$
for na\"ive fermions and one can extract the coefficient of
the $\sqrt{b}$ term using a tree level calculation.
The coefficient of the $\sqrt{b}$ term (equivalently, the
constant term in (\ref{aleqn})) will
depend on the fermion
discretization but one can show on general grounds that this
coefficient is positive for any discretization.
The doubly periodic function, $f(u,v)$, in (\ref{naivesingle})
is the 
na\"ive fermion determinant on a single site lattice in the
presence of a $U(1)$ gauge field of the form
$\left(e^{iu},e^{iv}\right)$ and
is a positive function for all $(u,v)$.
Different fermion discretizations will lead to different
expressions for $f(u,v)$ but they will all have the
general property that it is a positive function for
all $(u,v)$.
It follows that 
\be
\bar f (0,0) > \bar f(n,m) \ \ \ \ \ \ \ \ \forall (n,m)\ne (0,0)
\ee
in (\ref{ffour}).
Since $\bar f(n,m)$ are raised to the $N$th power upon
integration over the gauge fields in (\ref{treeg}) as
a consequence of
the identity in Appendix~\ref{intI}, it follows that
$\bar f^N(0,0)$ will dominate in the large $N$ limit
resulting in a positive constant term in (\ref{aleqn}) and
there by
driving the critical chemical potential to infinity.

The above argument shows that there is no critical
chemical potential for quarks in two dimensional large
$N$ QCD. 
If Eguchi-Kawai reduction is not valid, as is the case in $d>2$, then we
cannot work on a finite lattice in the continuum limit and
the arguments in this paper do not hold.

Two dimensional QCD has also been studied in the Hamiltonian
formalism~\cite{Salcedo:1990rw,Schon:2000qy}. Baryons have
been discussed in the Hartree-Fock approximation~\cite{Salcedo:1990rw}
and the role of chemical potential and temperature have also been 
discussed~\cite{Schon:2000qy}. Contrary to the 
Euclidean Lagrangian formalism where reduction 
makes the independence on temperature quite obvious,
it is harder to see the same result in the Hamiltonian 
formalism~\cite{Schon:2000qy}. Arguments in support of
the existence of a critical chemical potential is provided
in~\cite{Schon:2000qy} along with the possible breakdown of
translational invariance at high densities. 
The dependence on spatial momentum is suppressed in our
analysis since $M$ dominates the denominator of the integrand
in (\ref{npmu}). This can also be seen in the expression
for the critical chemical potential in (\ref{critchem})
and this is a consequence of the integration over
the Polyakov loop variables defined in (\ref{treeg}).
The Hamiltonian formalism,
by construction, is in the continuum time limit and starts with
the theory in the Weyl (temporal) gauge ($A_0=0$) and
imposes periodic boundary conditions on the spatial gauge field
in the time direction. This amounts to setting the Polyakov loop
in the time direction to unity which is not consistent
with the unbroken $Z_N$ symmetry.
Recently, the importance of the
role of Polyakov loop variables in the spatial direction
has been addressed 
in the Hamiltonian formalism~\cite{Bringoltz:2008iu}.

\acknowledgments

A.H. and R.N. acknowledges partial support by the NSF under grant number
PHY-055375. A.H. also acknowledges partial support by US Department of Energy grant under contract DE-FG02-01ER41172. R.N. would like to thank Barak Bringoltz for several useful discussions.

\appendix

\section{A useful integral}\label{intI}

Let $f(u,v)$ be a periodic function of
$u$ and $v$ with period $2\pi$. 
We decompose it into its fourier components according to
\be
f(u,v) = \sum_{n=-\infty}^\infty 
\sum_{m=-\infty}^\infty 
\bar f (n,m) e^{i \left(nu+mv\right)}.
\label{ffour}
\ee
Then, the integral
\be
I_N(p,q) 
= \int \prod_{i=1}^N \frac{du_i}{2\pi}
\prod_{i=1}^N \frac{dv_i}{2\pi}
\prod_{i=1}^N f\left(u_i+p,v_i+q\right)
2\pi \delta\left (\sum_{i=1}^Nu_i \right)
2\pi \delta\left (\sum_{i=1}^Nv_i \right)
\ee
is
\begin{eqnarray}
I_N(p,q)
&=& \sum_{n_i=-\infty}^\infty
\sum_{m_i=-\infty}^\infty
\left[ \prod_{i=1}^N \bar f(n_i,m_i) \right]\cr
&&\ \ \ 
\int \prod_{i=1}^N \frac{du_i}{2\pi}
\int \prod_{i=1}^N \frac{dv_i}{2\pi}
e^{i \sum_{i=1}^N 
n_i\left(u_i+p\right)
+m_i\left(v_i+q\right)
}2\pi \delta\left (\sum_{i=1}^Nu_i \right)
2\pi \delta\left (\sum_{i=1}^Nv_i \right)
\cr
&=& \sum_{n=-\infty}^\infty
\sum_{m=-\infty}^\infty
\bar f^N(n,m) e^{iN\left( n p + mq\right)}.
\end{eqnarray}

\end{document}